\documentclass[review]{elsarticle}
\usepackage{setspace}
\usepackage{amsmath}
\usepackage{graphics}

\begin{document}
%%\begin{spacing}{3.0}
%%\begin{spacing}{1.0}

\title{An analysis of the Brown-Biefeld effect}
\author{Reuven Ianconescu}
%\email{riancon@mail.shenkar.ac.il}
\ead{riancon@mail.shenkar.ac.il}
\author{Daniela Sohar}
\author{Moshe Mudrik}
%\affiliation{Shenkar College of Engineering and Design}
\address{Shenkar College of Engineering and Design}
%\date{}

\begin{abstract}
\noindent When a high voltage is applied on an asymmetric capacitor,
it experiences a force acting toward its thinner electrode. This
effect is called Brown-Biefeld effect (BB), after its discoverers
Thomas-Townsend Brown and Paul-Alfred Biefeld. Many theories have been
proposed to explain this effect, and many speculations can be found on
the net suggesting the effect is an antigravitation or a space warp
effect. However, in the recent years, more an more researchers
attribute the BB effect to a unicharge ion wind.  This work calculates
the levitation force due to ion wind and presents experimental results
which confirm the theoretical results.\\
\end{abstract}

\begin{keyword}
brown-biefeld, lifters, electrostatics, corona, ion drift, Deutsch assumption
\end{keyword}

\maketitle

\noindent PACS: 41.20.Cv, 52.30.-q, 41.20.-q\\
%41.20.Cv 	Electrostatics; Poisson and Laplace equations, boundary-value problems
%52.30.-q       Plasma dynamics and flow
%41.20.-q 	Applied classical electromagnetism
\\
\\

\newpage
\noindent{\bf\large 1. INTRODUCTION}\\

\noindent The Brown-Biefeld (BB) effect has been discovered in 1920 by
Thomas Townsend Brown and Paul Alfred Biefeld during their experiments
with Coolidge X-ray tube. They observed a thrust acting toward the thin
electrode. Thomas Townsend Brown made an extensive research on this effect
and wrote several patents \protect\cite{ttbrown1, ttbrown2, ttbrown3}.

There is a site dedicated to the BB effect \protect\cite{bb-web} and a site
dedicated to experiments of this effect \protect\cite{jln}, but very few
theoretical works have been written on the subject. 
Some of those tried to explain the effect by electro-gravitation
\protect\cite{tajmar-matos, takaaki-musha, montalk}
or thermodynamics \protect\cite{bahder-fazi}.

However, in the recent years more and more sources
\protect\cite{tajmar, zhao-adamiak, zhao-adamiak1, blazelabs}
attribute the effect to corona ionic air propulsion.  In
\protect\cite{blazelabs} a lot of experiments are described, and the
thrust force is described by approximate formulas based on ion
propulsion, and in \protect\cite{zhao-adamiak} a full calculation of
the thrust, based on the jet of the corona ion wind is performed.
The calculations in \protect\cite{zhao-adamiak} were very accurate, but
seemed to require a substantial amount of computing power.

In this work we also calculate the force due to ionic air propulsion,
but we adopt a different approach, based on the Deutsch assumption
\protect\cite{deutsch}, which has been extensively used in the
unipolar charge flow literature \protect\cite{tsyrlin, ieta, sigmond,
sigmond1, amoruso, popkov, sarmajan, bouziane1}, but also criticized (see J. E. Jones et al.
\protect\cite{jones} and A. Bouziane et al. \protect\cite{bouziane}),
by showing big differences in magnitude and direction
between the exact electric fields and the ones obtained with the
Deutsch assumption. However our goal is to calculate the thrust
force and for this purpose the Deutsch assumption proves useful.

The Deutsch assumption states that the equipotential surfaces of the
Laplacian problem are equipotential also for the Poissonian problem,
only with different values of potential. Of course under this
assumption, the electric field lines of the Laplacian and Poissonian
problems are in the same direction at any location.

As we shall see, the calculations based on Deutsch assumption
\protect\cite{deutsch}, within the Warburg region
\protect\cite{warburg1, warburg2} result in formulas
which fit very well the experiment.

In Section 2 we explain the operation principle and the configuration
which has been analyzed. We also present the equations that have
to be solved.

In Section 3 we present the basics of the Deutsch assumption. Sigmond
\protect\cite{sigmond1} made a profound analysis on the subject, and
summarized the work of many researchers concerning the use of the
Deutsch assumption. We will summarize those findings, and bring some
highlights on this issue.

In Section 4 we solve the Laplacian problem and calculate the
capacitance.  Those results are new, because as far as we know this
Laplacian configuration has not been solved yet.  Also, this Laplacian
solution is needed for the Poissonian problem when Deutsch assumption
is used.

In Section 5 we solve the Poissonian problem using the Laplacian
solution and Deutsch assumption and display the calculated force
and current.

In Section 6 we describe our experiment and show the obtained results.

In Section 7 we compare the measured and calculated results and express
our calculated results by approximate formulas.

The work is ended with some concluding remarks.\\

\noindent{\bf\large 2. THE OPERATION PRINCIPLE AND THE CONFIGURATION}\\

\begin{figure}[h]
\includegraphics[width=15cm]{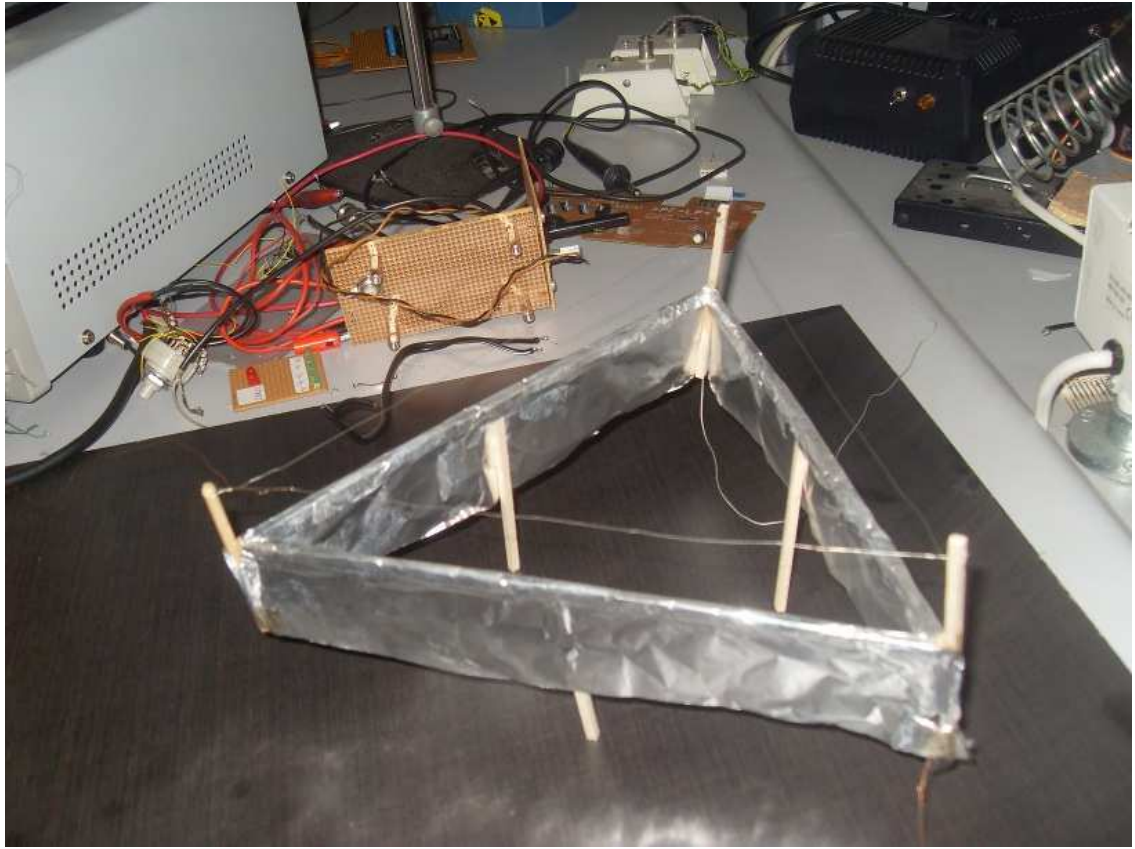}
%Created from lifter.gif, width=150
\caption{(color online) The picture of the lifter we built and used for experiments.
It consists of a thin anode wire over flat vertical cathode and is of
triangular form.}
\label{lifter}
\end{figure}

The picture of the lifter on which we did our experiments is shown in
Figure~\protect\ref{lifter}. The lifter is based on a thin anode
wire at positive high voltage, over a grounded flat cathode. As
follows from the explanation below, for best propulsion, the flat
electrode must be vertical. We made some
experiments with negative corona and the thrust is considerably
lower for negative corona than for positive corona, fact which
is confirmed by Blazelabs \protect\cite{blazelabs}. Hence we start
the explanation for positive corona.

The working principle is as follows: when a high enough positive
voltage is supplied, pairs of positive ion and electron (which are
always randomly created by incident photons) are accelerated: the
positive ion toward the cathode and the electron toward the anode.
Some of those pairs recombine emitting a photon which by the
photoelectric effect on a neutral atom, ionizes it, creating the
electron avalanche. High energy electrons hitting neutral atoms, ionize
them, contributing to the avalanche. This process arrives to a steady
state consisting in a corona region around the anode (in which the
ionization process happens) and in which there are positive
ions and high speed electrons.
Hence this region is neutral and typically narrow and outside
it there is a unicharge positive ion drift moving toward the
cathode. The positive ions transfer most of their momentum to neutral
molecules and while the positive ions ``feel'' the force of the
electric field and hence move toward the cathode (according to the
positive ion mobility coefficient) the neutral molecules keep the
inertia of the momentum they received. If the flat cathode were
horizontal (and infinite), the jet of neutral molecules would hit the
cathode and hence the forces on the anode and cathode would have been
equal and opposite, hence no net thrust.

However, the cathode being vertical, and given the fact that the ions
transfer {\it most} of their momentum to neutral molecules, those do not hit
the cathode, but form an air jet downstairs, and by momentum conservation
the lifter senses a net force upwards. Hence, the thrust force is calculated
as the {\it total force on the space charge}.

Also we treat the whole region as unicharge, and neglect the corona
thickness, which is small for the case of positive corona
\protect\cite{zhao-adamiak, blazelabs, peek}.

About negative corona: negative corona starts like the positive corona by
randomly created pairs of positive ions and electrons, only here the positive
ions are accelerated toward the negative thin electrode (called here cathode)
and the electrons are accelerated toward the flat electrode (called now anode).
Also for negative corona some pairs recombine, emitting a photon, but this time
the photoelectric effect created by it is on the thin wire surface, which being
negative, easily releases electrons, creating the electron avalanche. Also
electrons hit neutral atoms and ionize them, but because the electrons move
outward the negative thin electrode, this happens at lower velocities than for
positive corona, hence this part of the process is less dominant. Hence,
the photoelectric effect on the thin wire surface being dominant, the
negative corona is very sensitive to the ability of the thin wire to emit
electrons. If the thin wire surface has irregularities only some parts
of it emit electrons, creating tufts (see Peek \protect\cite{peek}, Fig.~77).
Also for a limiting voltage, for which the thin wire surface is not
able to steadily emit electrons, the process may be bursty.
This happens
because the positive ions attracted to the thin wire lower the strength
of the electric field near it, and this field is restored only after
the emitted electrons enter the flat positive conductor and arrive through
the power source to the thin negative wire. If the process arrives to
a steady state, the negative corona has two layers: the inner layer
called ionization layer which consists of positive ions and out flowing
electrons emitted by
the thin wire, the outer layer in which electrons flow out bouncing
into neutral atoms and combine in negative ions. Outside those layers
there is a unicharge flow of negative ions and electrons (see
\protect\cite{chen}, Fig.~1). We suppose
the main reasons for the levitating force being smaller with negative corona
are the tufts (reflecting a nonuniform ability of the thin wire to emit
electrons) and burstiness which may persist also after corona builds up.
We will not further deal with negative corona and all the
results obtained in this paper are valid for positive corona only.

\begin{figure}[h]
{\par\centering \includegraphics{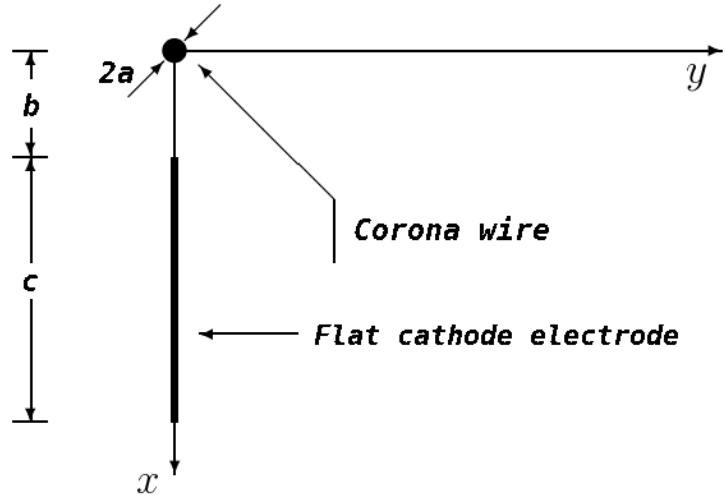} \par}
%Created from config.gif, width=100
\caption{The simplified 2D configuration. The corona wire is at coordinates
$(0,0)$ and its radius is $a$. The distance between the electrodes is called
$b-a$, and the cathode length is $c$.}
\label{simpconfig}
\end{figure}

Now the shape of the lifter must not necessarily be triangular, it
may be rectangular or any
other shape. Neglecting edge effects, the operation is described by a
thin anode wire, over a vertical conducting plane. Hence we deal with a
two dimensional problem described in Figure~\protect\ref{simpconfig}.

Our lifter has the following dimensions: $a=0.075\rm{mm}$, $b=2.8\rm{cm}$ and $c=4\rm{cm}$
(see Figure~\protect\ref{simpconfig}). Each side of the triangle is $0.2\rm{m}$,
hence if we calculate the thrust force
per unit length, we have to multiply by the perimeter of $3\times 0.2=0.6\rm{m}$
to find the total force.

As long as the potential difference is below the corona inception voltage
\protect\cite{peek}, there is no space charge, hence we have a {\it Laplacian}
problem, defined by 

\begin{equation}
\nabla^2 V_L=0,
\label{laplace}
\end{equation}

where the index ``L'' denotes the Laplacian solution. The boundary
conditions are $V_L=V_0$ on the anode wire surface and $V_L=0$ on the
cathode wire surface, where $V_0$ is the applied voltage. The electric
field is $\bar{E_L} = -\bar{\nabla} V_L$.

In presence of space charge, different kinds of ions have different
mobilities, defining the velocity of each ion as its mobility times
the electric field. However, we use the average mobility
\protect\cite{blazelabs, sigmond, jones} for positive air ions, known to be
$\mu=2\times 10^{-4} m^2/v\ sec$. Diffusion can usually be neglected
\protect\cite{felici, goldman}, hence the unipolar non diffusive drift of ions
is described by

\begin{equation}
\bar{J}= \rho \bar{v} = \mu \rho \bar{E_P},
\label{jrhoe}
\end{equation}

where $\rho$ is the space charge per unit of volume and $\bar{E_P}$ is not
the Laplacian field, but the {\it Poissonian} field, influenced by the
space charge itself via Gauss law:

\begin{equation}
\epsilon_0 \bar{\nabla} \cdot  \bar{E_p}= \rho.
\label{poisson}
\end{equation}

Being a stationary problem, the current conservation condition
$\bar{\nabla} \cdot \bar{J}=0$ must hold. So the Poissonian problem
may be formulated by: $\bar{\nabla} \cdot (\rho \bar{E_P})=0$ or

\begin{equation}
\bar{\nabla} \cdot (\nabla^2 V_P \ \bar{\nabla} V_P) = 0.
\label{third}
\end{equation}

Clearly, this equation contains the 3rd derivative, hence one needs an
additional boundary condition, and what is usually assumed is Kaptzov
hypothesis \protect\cite{kaptzov}.

Physically Kaptzov hypothesis means that once the potential difference has
been raised sufficiently for the corona to start, the electric field near
the corona conductor remains constant and equals the {\it inception} value,
even when further raising the potential difference.

Our goal is to calculate the thrust force, i.e. the $x$ component of the
total force on the space charge $\iiint E_{Px} \rho \rm{d}^3r$, where
$E_{Px}$ is the $x$ component of $E_P$, and the integral is on the whole
free space. Using eq.~(\protect\ref{jrhoe}), the total thrust force may be
expressed as

\begin{equation}
F=\frac{1}{\mu}\iiint J_{x} \rm{d}^3r
\label{tot_force}
\end{equation}

and for this goal we need a full solution of eq.~(\protect\ref{third}).
Felici \protect\cite{felici} analyzed equation (\protect\ref{third}) and
gave some particular solutions, and Feng \protect\cite{feng} calculated an
exact solution of eq.~(\protect\ref{third}) for concentric cylinders.

As mentioned before, eq.~(\protect\ref{third}) is easier to
solve by using the Deutsch assumption \protect\cite{deutsch} and this item
is discussed in the following section.\\

\noindent{\bf\large 3. ANALYSIS OF THE DEUTSCH ASSUMPTION}\\

The Deutsch \protect\cite{deutsch} assumption (DA) states that the
equipotential surfaces of the Laplacian problem, defined by
eq.~(\protect\ref{laplace}) are equipotential also for the Poissonian
problem, defined by eq.~(\protect\ref{third}), only with different values
of potential. This assumption is not valid {\it in
general}, hence its usage violates some laws of physics. Sigmond
\protect\cite{sigmond1} studied the consequences of using DA, and
showed that there are two general approaches in using DA. One approach
called T-type, keeps $\bar{\nabla} \times \bar{E_P}=0$, but violates
current conservation. The other approach called L-type,
satisfies the current conservation $\bar{\nabla} \cdot \bar{J}=0$
but violates the orthogonality between flux lines $\bar{E_P}$ and
potential surfaces (i.e. results in $\bar{\nabla} \times \bar{E_P}\neq 0$).

For more details, the reader is referred to \protect\cite{sigmond1},
formulas (13)-(39), but for convenience we show here the main highlights
of the issue. Dealing with a two dimensional problem we shall use
orthogonal coordinates $u,v$ describing the field direction
and equipotential surfaces, respectively (the equivalents of $l,\sigma$ in
\protect\cite{sigmond1}). Accordingly, the unit vectors will be noted as
$\widehat{u},\widehat{v}$.

The assumption that the equipotential surfaces of the
Laplacian problem remain equipotential in the Poissonian problem
implies that the Poissonian field $\bar{E_P}$ can be expressed as

\begin{equation}
\bar{E_P}=\theta \bar{E_L},
\label{ep_el}
\end{equation}

where $\theta$ is a {\it scalar function}. It is crucial for our
solution to have a correct current distribution, because from
eq.~(\protect\ref{tot_force}) we see that only a
correct current can result in a correct force, hence we use the
L-type described above. Requiring current conservation and using
eq.~(\protect\ref{jrhoe}) and (\protect\ref{ep_el}) we obtain

\begin{equation}
\bar{\nabla} \cdot \bar{J}=\bar{\nabla} \cdot (\mu\rho\theta \bar{E_L})=
\bar{\nabla}(\mu\rho\theta)\cdot\bar{E_L}+
(\mu\rho\theta)\bar{\nabla}\cdot\bar{E_L}=
\bar{\nabla}(\mu\rho\theta)\cdot\bar{E_L}=0,
\label{curr_cons}
\end{equation}

because $\bar{\nabla}\cdot\bar{E_L}=0$. Now $\bar{E_L}$ pointing in
the $\widehat{u}$ direction, and being non zero, implies:

\begin{equation}
\frac{\partial}{\partial u}(\mu\rho\theta)=0,
\label{mu_rho_theta}
\end{equation}

which means that $\mu\rho\theta$ must be a function of $v$ only:

\begin{equation}
\mu\rho\theta=K(v).
\label{mu_rho_theta1}
\end{equation}

Using again eq.~(\protect\ref{jrhoe}) and (\protect\ref{ep_el}) we get

\begin{equation}
\bar{J}=K(v)\bar{E_L}.
\label{J_Kv_E_L}
\end{equation}

This method has been carried out by Popkov \protect\cite{popkov}
and he showed that one may obtain the potential difference on each
field line as a function of the current density on the collector,
and used cross-plot technique to obtain the current density in terms
of the applied voltage.
Sarma and Janischewskyj \protect\cite{sarmajan} followed the same
technique as Popkov, but used an iterative procedure to calculate
the current density on each field line in terms of the applied
voltage. A. Bouziane et al. \protect\cite{bouziane1}) used the
above methods to evaluate the current density and field
profile for several geometries, and found the results in agreement
with experiment.

Our goal is to calculate the thrust force in eq.~(\protect\ref{tot_force}),
hence we will not follow the above techniques, but
rather look for an adequate function $K(v)$. It will be helpful for
example if we could choose this function to be constant
in some regions of $v$.

Now because we used the L-type, $\bar{\nabla}\times\bar{E_P}\neq 0$
in general, so if we require $\bar{\nabla}\times\bar{E_P}=0$ we end
with a contradiction.

As we shall see in the next section, the Laplacian field lines are
almost straight in the paraxial region around $y=0$ (see
Figure~\protect\ref{simpconfig}). It is
therefore of interest to examine what happens
in a region where the Laplacian field lines $E_L$ are straight. In
such region the unit vector $\widehat{u}$ does not change with $u$, so
is a function $v$ only. Knowing that $\bar{\nabla}\times\bar{E_L}=0$,
results in $E_L$ being a function of $u$ only.

We shall see that in this case we may require
$\bar{\nabla}\times\bar{E_P}=0$, because this belongs to the cases
described in \protect\cite{sigmond1}, eqn. (38) and (39), for which
the DA holds exactly. So requiring:

\begin{equation}
0=\bar{\nabla}\times\bar{E_P}=\bar{\nabla}\times(\theta \bar{E_L})=
\bar{\nabla}\theta\times\bar{E_L} + \theta\,\,\bar{\nabla}\times\bar{E_L}=
\bar{\nabla}\theta\times\bar{E_L}
\label{curlEP0}
\end{equation}

results in
$\bar{\nabla}\theta$ being in the $\widehat{u}$ direction, hence
$\theta$ being only a function of $u$. Therefore
$\bar{E_P}=\theta\bar{E_L}$ is also a function of $u$ only, and
the same is true for $\rho=\epsilon_0\bar{\nabla}\cdot\bar{E_P}$.

The outcome is that $\mu\rho\theta$ in eq.~(\protect\ref{mu_rho_theta1})
which is {\it only} a function of $v$ must also be {\it only} a
function of $u$.  This is possible only if the function $K(v)$
in eq.~(\protect\ref{J_Kv_E_L}) is a constant.

This is helpful, since we may start our Poissonian calculations
(Sect.~5) in the paraxial region with $K(v)$ constant, and let it
drop to 0 when approaching the Warburg
\protect\cite{warburg1, warburg2} limit region.

This is in accordance with the experimental knowledge that the current
drops to 0 outside the Warburg region \protect\cite{kondo, selim1,
selim2, selim3} and also with Ieta \protect\cite{ieta}, who showed
that the Laplacian solution reconstructs well the Warburg distribution
for a pin to plane geometry. As we shall see, this method results in
a calculated thrust force which fits experiment.

So, for using Deutsch assumption in the Poissonian problem, we need the
solution to the Laplacian problem first, and this is done in the next
section.\\

\noindent{\bf\large 4. SOLUTION OF THE LAPLACIAN PROBLEM}\\

There are many ways to solve the Laplacian problem (analytic or numeric)
and we chose to solve it analytically, by separation of variables in cylindrical
coordinates.

The schematic configuration in Figure~\protect\ref{simpconfig}, does not
allow to properly define boundary conditions in separate variables in cylindrical
coordinates, hence we will make a slight change to this configuration.

Anyhow the configuration in Figure~\protect\ref{simpconfig} is not accurate,
because it does not show the width of the flat cathode. In practice the cathode
must have a finite width, and more than that, the curvature radius of the cathode
at the location $(x=b, y=0)$ (see Figure~\protect\ref{simpconfig}) must be big enough so that {\it no corona} can be
formed there \protect\cite{peek}.

\begin{figure}[h]
{\par\centering \includegraphics{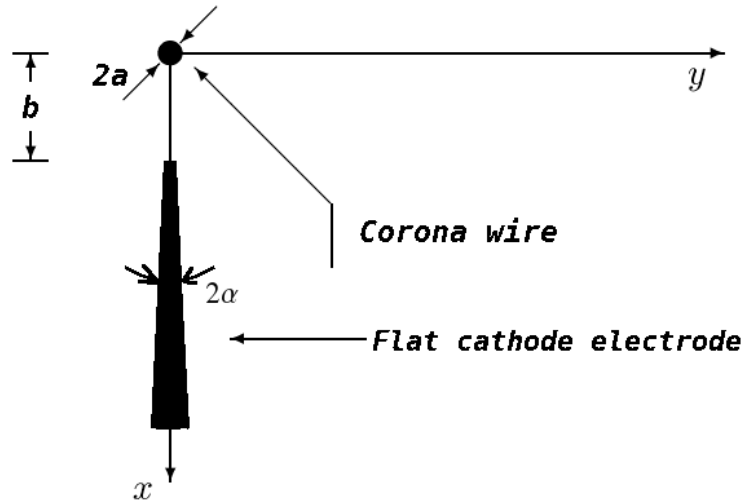} \par}
%Created from adapted_config.gif, width=100
\caption{Configuration adapted to cylindrical coordinates. The cathode
contour is described by the lines $r=b$, and $\varphi=\pm\alpha$.}
\label{modconfig}
\end{figure}

The modified configuration is described in Figure~\protect\ref{modconfig}.

The flat cathode surface is now defined along $\varphi=\pm\alpha$, where
$\alpha$ is a fixed small angle of about $1.3^o$ (based on the thickness
of the cathode in our lifter). This model still includes edges at
$(r=b,\varphi=\pm\alpha)$, but those edges do not exist in the real
engine and except of them the model is quite close to reality.
At last, we take the
cathode length $c$ to infinity: the main interaction is between the electrodes
and using a finite length $c$ would still be solvable analytically, but
would be an unnecessary complication.

We define the potential $V_1$ for the region $a\le r\le b$ and
$0\le\varphi\le 2\pi$ and the potential $V_2$ for the region $r\ge b$
and $\alpha\le\varphi\le 2\pi-\alpha$. The boundary conditions are:

\begin{equation}
V_1(r=a,\varphi)=V_0,
\label{boundary1reqa}
\end{equation}

where $V_0$ is the applied voltage. Because of the mirror symmetry
around the $x$ axis, we require

\begin{equation}
V_1(r,\varphi)=V_1(r,2\pi-\varphi).
\label{boundary1phieq0}
\end{equation}

The potential continuity at $r=b$ gives

\begin{equation}
V_1(r=b,\varphi)=\left\{
\begin{array}{ll}
0 & 0\le\varphi\le\alpha \\
0 & 2\pi-\alpha\le\varphi\le 2\pi \\
V_2(r=b,\varphi) & \alpha\le\varphi\le 2\pi-\alpha \\
\end{array}
\right.
\label{pot_cont_b}
\end{equation}

and the normal field continuity at $r=b$ requires

\begin{equation}
\partial_r V_1(r=b,\alpha\le\varphi\le 2\pi-\alpha)=\partial_r V_2(r=b,\alpha\le\varphi\le 2\pi-\alpha).
\label{boundary12dercont}
\end{equation}

The potential is $0$ on the sides of the big conductor

\begin{equation}
V_2(r,\varphi=\alpha)=V_2(r,\varphi=2\pi-\alpha)=0
\label{boundary2phialpha}
\end{equation}

and must go to $0$ at $r\rightarrow\infty$ :

\begin{equation}
V_2(r\rightarrow\infty,\alpha\le\varphi\le 2\pi-\alpha)=0.
\label{boundary2r_inft}
\end{equation}

We will use the well known solutions for the Laplace equation in
cylindrical coordinates, for the $z$ independent case, given by
the trivial solution $D-E\ln r$ (where $D$ and $E$ are constants)
plus the non trivial solution:

\begin{equation}
\sum_{\nu} (A_{\nu} r^{\nu}+B_{\nu} r^{-\nu})(C_{\nu}\cos(\nu\varphi)+D_{\nu}\sin(\nu\varphi)),
\label{non-trivial}
\end{equation}

where one may consider only non negative values of $\nu$, because negative
$\nu$ just switches the roles of $A_{\nu}$ and $B_{\nu}$.

Let us start with $V_2$. Condition (\protect\ref{boundary2r_inft})
excludes the trivial solution and the $r^{\nu}$ solution which diverge
at $r\rightarrow\infty$. Also, the requirement $V_2(r,\varphi=\alpha)=0$
in eq.~(\protect\ref{boundary2phialpha}) imposes a combination between the
$\sin$ and $\cos$ terms in eq.~(\protect\ref{non-trivial}) of the form
$\sin(\nu(\varphi-\alpha))$ and the constant $B_{\nu}$ may be normalized
for convenience to $B_{\nu}/b^{-\nu}$. Hence we may write the following
expression for $V_2$:

\begin{equation}
V_2=\sum_{\nu} B_{\nu} (r/b)^{-\nu}\sin(\nu(\varphi-\alpha)).
\label{expression1_V2}
\end{equation}

To satisfy the requirement $V_2(r,\varphi=2\pi-\alpha)=0$ in
eq.~(\protect\ref{boundary2phialpha}), we need
$\sin(\nu(2\pi-\alpha-\alpha))=0$, or $\nu(2\pi-2\alpha)=m\pi$ (where
$m$ is a positive integer), thus giving the values of
$\nu=\frac{m}{2}\frac{1}{1-\alpha/\pi}$. So the expression for $V_2$
may be written as:

\begin{equation}
V_2=\sum_{m=1}^{\infty} B_m (r/b)^{-\frac{m}{2}\frac{1}{1-\alpha/\pi}}\sin\left(\frac{m}{2}\frac{\varphi-\alpha}{1-\alpha/\pi}\right).
\label{expression2_V2}
\end{equation}

Now we look for an expression for $V_1$. Because the non trivial solution
is $\varphi$ dependent for any $r$, to satisfy condition
(\protect\ref{boundary1reqa}), we need also the trivial solution.
Also, because $V_1$ is defined in a region with circular continuity, adding
$2\pi$ to $\varphi$ must result in the same potential, hence $\nu$ must be
an integer, say $m$. For satisfying eq.~(\protect\ref{boundary1phieq0}) only
the $\cos$ solution must be taken, and again, we may normalize the constants
so that the power is on $r/a$ instead of $r$, so we may write the following
expression for $V_1$:

\begin{equation}
V_1=D-E\ln r+\sum_{m=0}^{\infty} \left(A_m (r/a)^m + C_m (r/a)^{-m}\right) \cos(m\varphi).
\label{expression1_V1}
\end{equation}

To satisfy condition (\protect\ref{boundary1reqa}), the $m\neq 0$ terms of
the non trivial part of eq.~(\protect\ref{expression1_V1}), must be identically
$0$ for $r=a$. Given that the functions $\cos(m\varphi)$ are orthogonal in
the interval $0\le\varphi\le 2\pi$, each term of the series must vanish
for $r=a$, hence we get $A_m=-C_m$, for $m\neq 0$. The $m=0$ term gives just
a constant, which may be absorbed in the trivial solution, but it proves
convenient to name the $m=0$ term $L_0$, and to scale separately the
trivial solution so that it results in an arbitrary constant $V'$ for
$r=a$, and $0$ for $r=b$, so that
$D-E\ln r=V'\left(1-\frac{\ln(r/a)}{\ln(b/a)}\right)$. So we obtain
for $V_1$:

\begin{equation}
V_1=V'\left(1-\frac{\ln(r/a)}{\ln(b/a)}\right)+L_0+\sum_{m=1}^{\infty} A_m \left( (r/a)^m - (r/a)^{-m}\right) \cos(m\varphi).
\label{expression2_V1}
\end{equation}

It is to be mentioned that we do not lose any generality with the
above scaling of the trivial solution, because after scaling, the
trivial solution plus $L_0$ result in $V'+L_0$ for $r=a$ and
$L_0$ for $r=b$, and the relation between them has not been
established yet.

The reason for choosing this approach is that all the unknowns, namely
$A_m$ and $L_0$ in eq.~(\protect\ref{expression2_V1}) and $B_m$ in
eq.~(\protect\ref{expression2_V2}) must be proportional to the applied
voltage $V_0$, so we may calculate them with the aid of an arbitrary
$V'$, and obtain $V_1(r=a,\varphi)=V'+L_0$, which can be scaled
eventually by a factor $V_0/(V'+L_0)$ to be equal to $V_0$. So we
may set from now on $V' \equiv 1$, and remember to multiply everything by
$V_0/(1+L_0)$.
One may verify that the alternative approach of absorbing $L_0$ in the
trivial solution results in much more complicated equations.

Here one has to require the boundary conditions and solve 2 sets
of matrix equations to find the vectors $A_m$, $B_m$ and $L_0$.
The details are worked out in the Appendix, and
Figure~\protect\ref{3D_potential} shows the potential in a 3D plot, for
the physical values of our lifter, where $V_0$ is normalized to 1.
Because of the mirror symmetry around the
$x$ axis, we drew the potential only for positive $y$.\\

\begin{figure}[h]
{\par\centering \includegraphics{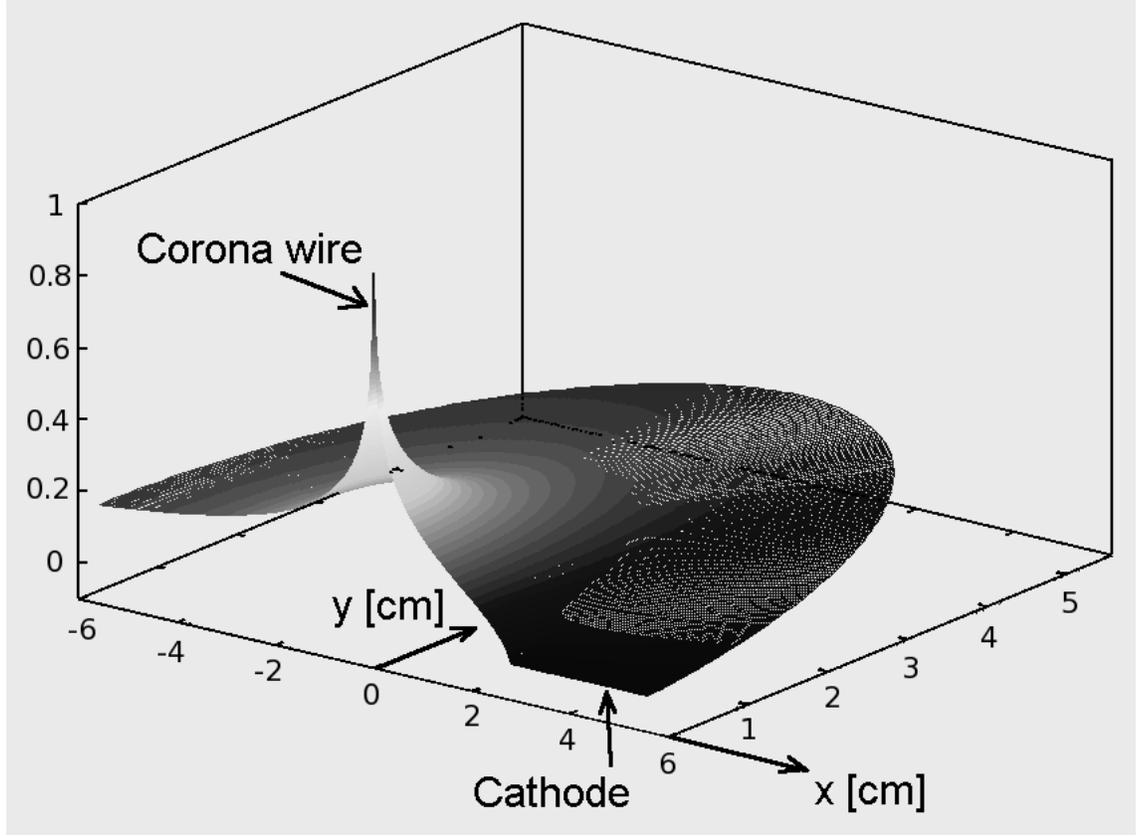} \par}
%Created from fig3.gif, width=150
\caption{3D plot for the calculated potential. The applied voltage
$V_0$ is normalized to 1. The calculations used the physical values
of our lifter, i.e. corona wire radius $a=0.075\rm{mm}$, distance
between electrodes $b=2.8\rm{cm}$ and the angle $\alpha=1.3^o$.}
\label{3D_potential}
\end{figure}

We may derive now the electric field. For region 1 we obtain the radial field:

\begin{equation}
E_{1r}= -\frac{\partial V_1}{\partial r}=\frac{V_0}{1+L_0}\frac{1}{r}\left[\frac{1}{\ln(b/a)}-\sum_{m=1}^{\infty} m A_m \left( (r/a)^m + (r/a)^{-m}\right) \cos(m\varphi) \right]
\label{expression1_E1r}
\end{equation}

and the circular field:

\begin{equation}
E_{1\varphi}= -\frac{1}{r}\frac{\partial V_1}{\partial \varphi}=\frac{V_0}{1+L_0}\frac{1}{r} \sum_{m=1}^{\infty} m A_m \left( (r/a)^m - (r/a)^{-m}\right) \sin(m\varphi).
\label{expression1_E1phi}
\end{equation}

For region 2 we obtain the radial field:

\begin{equation}
E_{2r}= -\frac{\partial V_2}{\partial r}=\frac{V_0}{1+L_0}\frac{1}{r}\sum_{l=1}^{\infty} \frac{l-1/2}{1-\alpha/\pi}  B_l (r/b)^{-\frac{l-1/2}{1-\alpha/\pi}}\sin\left((l-1/2)\frac{\varphi-\alpha}{1-\alpha/\pi}\right)
\label{expression_E2r}
\end{equation}

and the circular field:

\begin{equation}
E_{2\varphi}= -\frac{1}{r}\frac{\partial V_2}{\partial \varphi}=-\frac{V_0}{1+L_0}\frac{1}{r}\sum_{l=1}^{\infty} \frac{l-1/2}{1-\alpha/\pi}  B_l (r/b)^{-\frac{l-1/2}{1-\alpha/\pi}}\cos\left((l-1/2)\frac{\varphi-\alpha}{1-\alpha/\pi}\right).
\label{expression_E2phi}
\end{equation}

It would be useful to calculate the capacitance. The charge per unit of surface on the
anode is $\eta=\epsilon_0 E_{1r}(r=a,\varphi)$, and the charge per unit of length is given
by $\lambda=\int_0^{2\pi} a\ d\varphi\ \eta$. Clearly, the integral on $\varphi$ zeroes
the sum in eq.~(\protect\ref{expression1_E1r}) and we are left with:

\begin{equation}
\lambda = \frac{2\pi\epsilon_0 V_0}{\ln(b/a)(1+L_0)}.
\label{lambda}
\end{equation}

So the capacitance per unit of length is

\begin{equation}
C'=\frac{\lambda}{V_0} = \frac{2\pi\epsilon_0}{\ln(b/a)(1+L_0)}.
\label{capacitance}
\end{equation}

Of course, for $\alpha=\pi$, from eq.~(\protect\ref{expression1_L0})
we have $L_0=0$  and
we recover the known formula for the capacitance per unit of length for
concentric cylinders.

\begin{figure}[h]
{\par\centering \includegraphics{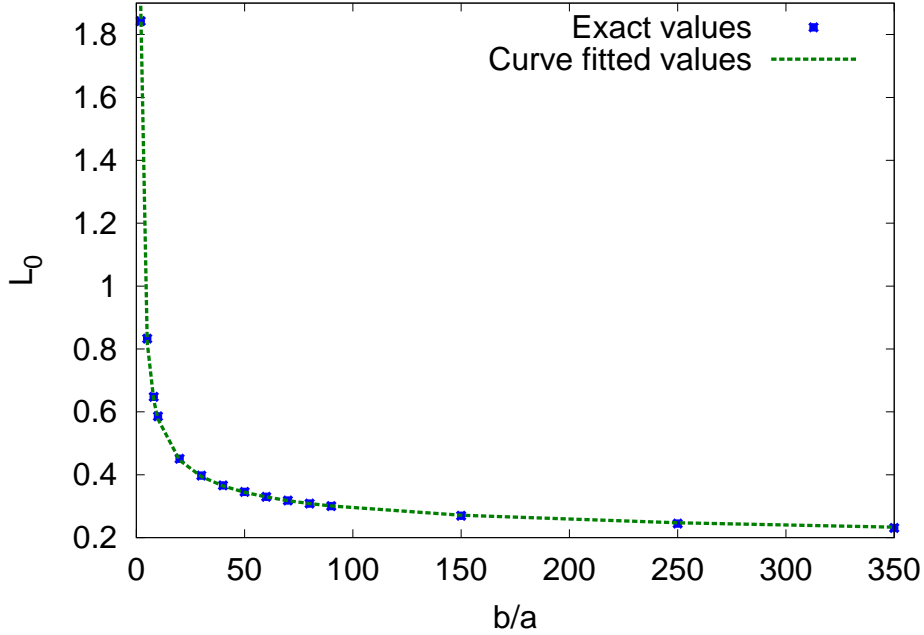} \par}
%Created from L0_curve_fitting.gif, width=150
\caption{(color online) Values of $L_0$ for $\alpha=1.3^o$. The stars are the
exact calculated values, and the continuous line represents the fitted curve
values given by eq.~\protect\ref{approx_L0}. This fitted curve is valid for
$\alpha<\pi/100\ \rm{rd}=1.8^o$.}
\label{L0}
\end{figure}

We are of course interested in small $\alpha$, so we calculated the
values of $L_0$ for different ratios $b/a$, and fit an approximate formula
for it - see Figure~\protect\ref{L0}.

It comes out that for $\alpha < \pi/100$, $L_0$ can
be expressed as

\begin{equation}
L_0 \approx \frac{1.3035}{\ln(b/a)} + 0.011,
\label{approx_L0}
\end{equation}

resulting in the following approximate formula for the capacitance per unit of
length:

\begin{equation}
C'\approx \frac{2\pi\epsilon_0}{1.011\ \ln(b/a)+1.3035}.
\label{approxcapacitance}
\end{equation}

If we set the values for our lifter: $a=0.075\rm{mm}$ and $b=2.8\rm{cm}$ into
eq.~(\protect\ref{approxcapacitance}) we get $C'=7.63\rm{pF/m}$, and multiplying by the lifter's perimeter $0.6\rm{m}$,
the calculated capacitance comes out $C_{\rm calculated}=4.57\rm{pF}$. To check the validity
of this result we measured the capacitance of our lifter and got
$C_{\rm measured}=4\rm{pF}\pm 5\%$, which is quite close to the calculated result.

It is also useful to get a relation between the applied voltage $V_0$ and the
field intensity on the anode. Of course, the field on the anode wire is not
constant, and depends on $\varphi$. But for a thin anode ($b \gg a$), the
field is almost constant (see also discussion in the next section).

One may verify that for $r=a$ the absolute value of the sum in
eq.~(\protect\ref{expression1_E1r}) is much smaller than $\frac{1}{\ln(b/a)}$
hence we may write the approximate expression:

\begin{equation}
\frac{V_0}{E_{1r}(r=a,\varphi)}\approx {(1+L_0)\ a \ln(b/a)} \approx a(1.3035+1.011\ln(b/a)),
\label{approxVoverE}
\end{equation}

where the second expression is a further approximation which uses
eq.~(\protect\ref{approx_L0}).

Again, for $\alpha=\pi$, $L_0=0$, and we recover the known expression
of this relation for concentric cylinders.\\

\begin{figure}[h]
{\par\centering \includegraphics{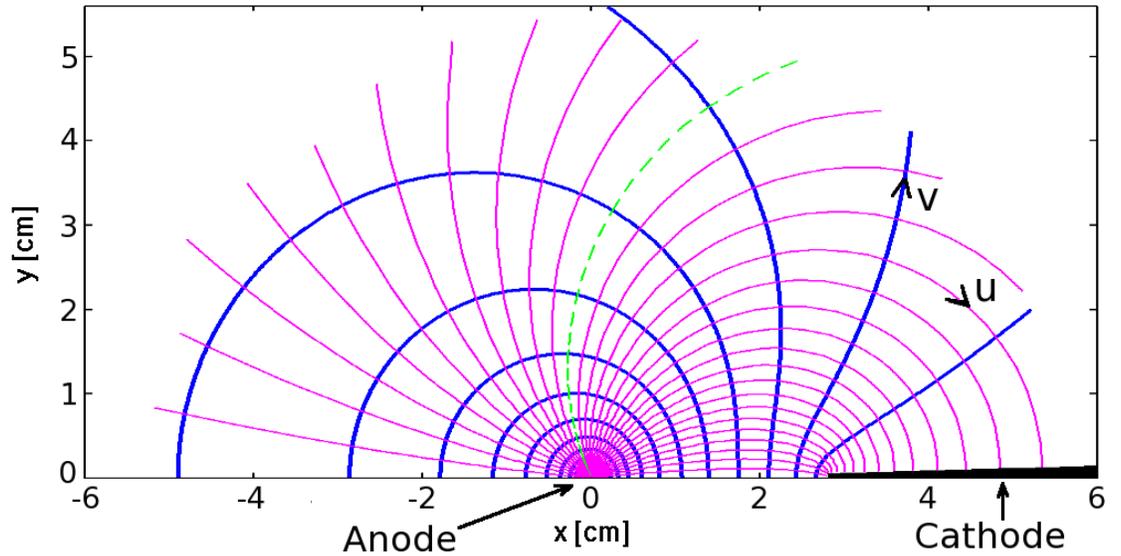} \par}
%Created from equipotential_and_field_lines.gif, width=150
\caption{(color online) The thin lines emerging from the
anode are the electric field lines while the thick lines are
the equipotential surfaces. The coordinates $u$ and $v$ denote the
local direction of the field and the equipotential lines,
respectively. The dashed line represents the electric field
line which delimits the Warburg region, hence passes through the point
$(x=b, y=b/\tan(60^o))$. Only the $y>0$ region is shown, because of
the mirror symmetry around the $x$ axis.}
\label{equipot}
\end{figure}

In Figure~\protect\ref{equipot} we show the equipotential surfaces and the
field lines - those are needed in the next section.

The Warburg
\protect\cite{warburg1, warburg2} region can
be seen also in Figure~\protect\ref{equipot}, and will be referred to
in the next section.\\

\noindent{\bf\large 5. THE POISSONIAN PROBLEM}\\

In this section we use the Laplacian results obtained in the previous section
to solve the Poissonian problem, i.e. the state of the system when the applied
voltage is bigger than the corona inception voltage.

F.W. Peek \protect\cite{peek} made an extensive research on corona inception
for different geometries like concentric cylinders, parallel wires, etc.
and published
the results in his book. For all the configurations
involving corona around a thin wire of radius $a$, the electric field on
the surface of the wire at which corona begins (at room temperature)
is given by $E_i = 3\times 10^6 (1+p/\sqrt{a})\ V/m$, where $p\approx 0.03\ \sqrt{m}$
with {\it very slight} variations of about $2\%$ for different geometries,
with different asymmetries for the electric field (as found by Peek
\protect\cite{peek}, page 63).

We do not have the exact value for our geometry, but as explained by
Peek himself, before corona starts, the field very close to a thin
wire behaves like the field on the surface (which is almost constant
if the wire is thin) times the wire radius, divided by the distance
from the wire. So when having the above $E_i$ value {\it on the wire
surface}, one can easily find that at a distance of $p \sqrt{a}$ from
the wire surface the field is $3\times 10^6\ V/m$, this way assuring a
field of more than $3\times 10^6\ V/m$ in a wide enough region around
the wire to allow corona to start.

So we can safely use Peek formula:

\begin{equation}
E_i = 3\times 10^6 \left(1+0.03/\sqrt{a}\right)\ V/m
\label{peekformula}
\end{equation}

as has been done a lot in the literature, for different configurations
of thin wire electrode near any other electrode
\protect\cite{zhao-adamiak, jones, feng}.

Given $a=0.075mm$, we know that for our case $E_i=13.392\ MV/m$.

Now we can find at which voltage the corona starts, i.e. the Corona
Inception Voltage (CIV). The CIV is the
voltage for which the Laplacian field on the surface of the corona
wire equals to $E_i$. For this we do not have to use Peek formula
for CIV, we have the Laplacian solution for our problem. Using the approximation
(\protect\ref{approxVoverE}), results in $7.32\ KV$ or running the solution
and measuring the exact relation results in $7.42\ KV$, so the difference is
less than $2\%$. Hence we may use:

\begin{equation}
{\rm CIV}=7.4\ KV.
\label{civ}
\end{equation}

Because of the asymmetry around the anode, the field on the corona
surface is not completely uniform, and that is the reason for the
slight variations in $p$ in Peek formula for different asymmetric
configurations. But fortunately, the corona wire being very thin, the
field on the corona wire surface is almost uniform (up to variations
of $0.5\%$), so we do not have to worry about it. (See also discussion
before eq.~(\protect\ref{approxVoverE})).

So for a voltage bigger than the CIV, the Laplacian solution is not
valid anymore, and we need the solution to the Poissonian problem.
This requires the simultaneous solution of eq.~(\protect\ref{poisson}) and
(\protect\ref{jrhoe}), where for $J$ we use eq.~(\protect\ref{J_Kv_E_L}).

The coefficient $K$ in eq.~(\protect\ref{J_Kv_E_L}) is unknown, but given
the fact that the Deutsch assumption is accurate for $y=0$,
i.e. where the field lines are straight (see discussion in Sect.~3 and
eqn.~(38) and (39) in Sigmond \protect\cite{sigmond1}),
one can iterate the coefficient $K$ in this region to get $\int_a^b \! E \, \mathrm{d}x = V_0$, where $V_0$ is the potential difference
for which we solve. 

\begin{figure}[h]
{\par\centering \includegraphics{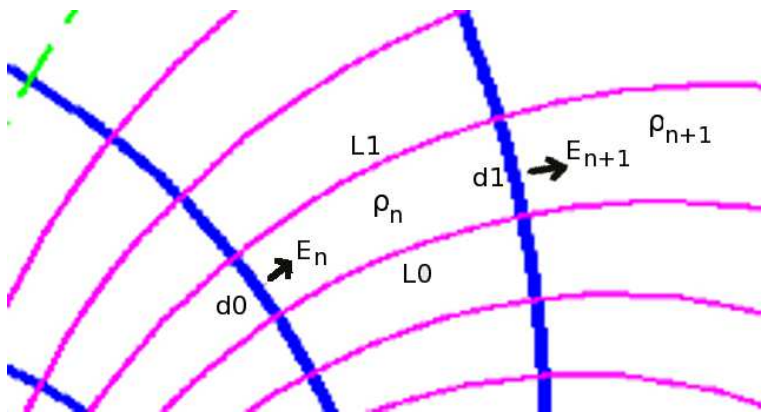} \par}
%Created from numerical_solution.gif, width=100
\caption{(color online) Zoom on a region Figure~\protect\ref{equipot}. The calculations are
done along the Laplacian field lines, and start with $n=0$ on the anode
surface up the final value of $n$ on the cathode surface.  The lengths on
the sides of each area are called $d0$, $d1$, $L0$ and $L1$.}
\label{numsol}
\end{figure}

The numerical solution is described in Figure~\protect\ref{numsol}, which is
a zoom on a region of Figure~\protect\ref{equipot}.

The ``P'' prefix for Poissonian is omitted for brevity, and having
a two dimensional problem $J$ represents the current per unit of
perpendicular length, and $\rho$ represents the charge per unit of
surface. The iterative calculation is done between each pair of
field lines (see Figure~\protect\ref{numsol}) by:

\begin{equation}
\rho_n=J_n/(\mu E_n),
\label{jrhoe_num}
\end{equation}

where $J_n$ is $J$ at the location of $E_n$, and it is known for a given
$K$, and

\begin{equation}
E_{n+1}=(\rho_n d_{av} L_{av}/\epsilon_0 + d0\ E_n)/d1,
\label{poisson_num}
\end{equation}

where $d_{av}=(d0+d1)/2$ and $L_{av}=(L0+L1)/2$. Those are the
numerical implementations of eq.~(\protect\ref{jrhoe}) and
(\protect\ref{poisson}).
The initial condition for the iteration is Kaptzov assumption
\protect\cite{kaptzov} (explained in the introduction), which requires:

\begin{equation}
E_0=E_i.
\label{kaptzov_num}
\end{equation}

After finishing the calculation between the first two field lines (i.e.
in the region of very small $y$ we check the result of
$\int_a^b \! E \, \mathrm{d}x \approx \displaystyle\sum\limits_n E_n L_{av_n}$.
Say its value is $2 V_0$, we
have to reduce $K$ by a factor of approximately 4 (approximate because
the initial condition for $E$ is independent on $K$). This process
converges very quickly (3-4 iterations), and after establishing $K$
we can process the calculations.

As explained in Sect.~3, we let $K$ drop to 0 when reaching the
Warburg \protect\cite{warburg1, warburg2} limit region
(see Figure~\protect\ref{equipot}).
Within each area element we also calculate the $x$ component of the force

\begin{equation}
F_{x_n}=E_{x_n} \rho_n ,
\label{fx}
\end{equation}

where $E_{x_n}$ is the $x$ component of $E_n$. The total force is eventually
summed on the whole area, and knowing $J$ we sum across the line fields
(in the $v$ direction), obtaining the total current $I$.

In the final stage the force and the current are multiplied by 2 to account
for the symmetric $y<0$ region and the values being per unit of length of
the lifter, are multiplied by the perimeter $0.6\rm{m}$. The force is normalized
to show the lifted mass in grams.

The values of $V_0$ for which we did the
calculations, have been chosen to correspond to the values on which the
experiment has been done (see next section).
The calculated results are presented in Table~\protect\ref{calc}.

\begin{table}[htbp!]
\caption{Calculated results}
\centering
\begin{tabular}{| l | l | l | }
\hline\hline
   $V_0\ [KV]$ &  Current [mA] &  Mass that can be lifted [g] \\
\hline
   11.12  &    0.084 &  1.54      \\
   12.9   &    0.145 &  2.65      \\
   14.08  &    0.19  &  3.55      \\
   14.4   &    0.219 &  4         \\
   15.64  &    0.27  &  4.98      \\
   16.8   &    0.34  &  6.16      \\
   18.9   &    0.46  &  8.42      \\
   20.5   &    0.586 &  10.74     \\
   21.8   &    0.678 &  12.44     \\
   22.8   &    0.76  &  13.9      \\
\hline  
\end{tabular}\\
\label{calc}
\end{table}

\noindent{\bf\large 6. THE EXPERIMENT}\\

\begin{figure}[h]
{\par\centering \includegraphics{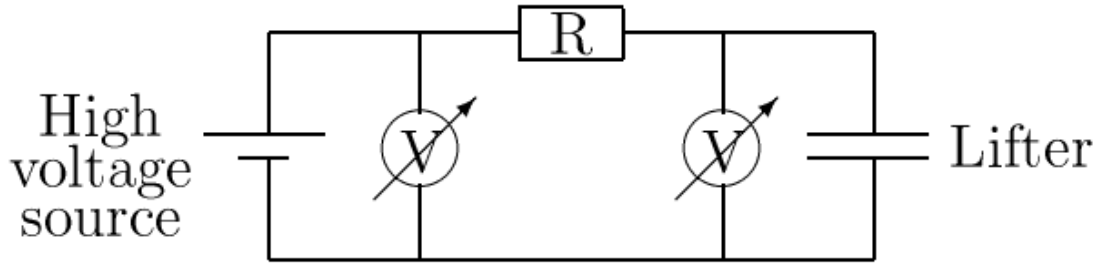} \par}
%Created from experiment_configuration.gif, width=150
\caption{The diagram of the experimental setup. The lifter is connected to a
high voltage source through a resistor. For different values of resistors,
the input and lifter voltages are measured. The lifting force is measured
in the cases it was big enough to lift.}
\label{experiment}
\end{figure}

The diagram of the experimental setup is shown in
Figure~\protect\ref{experiment}.

We used the lifter shown in Figure~\protect\ref{lifter} for our
experiments. The corona wire is a regular copper wire of radius
$0.075\rm{mm}$, and the cathode is an aluminum foil of width
$4\rm{cm}$ (called $a$ and $c$ in Figure~\protect\ref{simpconfig},
respectively). The sticks that hold the device are of balsa wood,
and the perimeter of the engine is $60\rm{cm}$. The above
materials are relatively light, so that the mass of the lifter 
is $7\rm{g}$.

For different values of resistors, the input and lifter voltages are measured
and the lifting force is measured. The lifting force has been measured only
for the cases for which the lifter lifted, by counter balancing it, knowing
that its mass is $7\rm{g}$.

The results of the measurements are shown Table~\protect\ref{meas}. The
current is calculated using the 2 measured voltages and the resistor.

\begin{table}[htbp!]
\caption{Measured results (NA means not available)}
\centering
\begin{tabular}{| l | l | l | l | l |}
\hline\hline
   $R\ [M\Omega]$ & Lifter voltage $V_0$ [KV] & Source voltage [KV] & Current [mA] &  Lifted mass [g] \\
\hline
308  &    11.12  &    25.6  &  0.047     &    NA    \\
154  &    12.9   &    25.58 &  0.082     &    NA    \\
110  &    14.08  &    25.51 &  0.104     &    NA    \\
88   &    14.4   &    25.46 &  0.126     &    NA    \\
66   &    15.64  &    25.42 &  0.148     &    NA    \\
44   &    16.8   &    25.3  &  0.193     &    NA    \\
22   &    18.9   &    25    &  0.277     &   7.5    \\
10   &    20.5   &    25    &  0.450     &   10	    \\
6.8  &    21.8   &    24.92 &  0.459     &   12.2   \\
3.3  &    22.8   &    24.8  &  0.606     &   13	    \\
\hline  
\end{tabular}\\
\label{meas}
\end{table}

\noindent{\bf\large 7. COMPARISON AND APPROXIMATED FORMULAS}\\

One may see that the calculated forces fit well to the measured forces,
with an average deviation of about 6.5\%. This suggests that our method gives a
good estimate for the lifting force.

Let us first analyze the relation between force and current, by
comparing it with a simpler case of straight, parallel and uniform field lines.
If the field lines are in the $x$ direction, our coordinates $u,v$ (see
Figure~\protect\ref{equipot}) correspond to $x,y$ respectively. For this
case, applying eq.~(\protect\ref{tot_force})
$F=\frac{1}{\mu}\iiint J_{x} \rm{d}^3r$ and integrating over the $y$ and
$z$ coordinates yields $F=\frac{I}{\mu}\int \rm{d}x$. The integral just
gives the length of the field lines, let us call it $d$, hence
the force {\it on the ionic space charge} is $F=(I d/\mu)$.

Certainly, if this configuration is a parallel plates capacitor,
the above force on the space charge does not produce thrust, because the
wind hits one plate. But one may build a lifter with approximate
straight, parallel and uniform field lines by using many thin wires on
top of many flat vertical aluminum foil cathodes (see
\protect\cite{blazelabs}). And for such a lifter $F=(I d/\mu)$ describes
well the thrust force.

Examining the connection between our calculated force and current in
Table~\protect\ref{calc} we find that the relation between force and
current is $18.4 \ \rm{g/mA} \pm 0.8\%$, or in terms of force instead of
mass, we get $179.72 \ \rm{N/A}$. As explained above for parallel field,
we would expect this relation to be proportional to $b$, which is the
distance between the electrodes in our lifter.

Hence we repeated the procedure explained in Sect.~5 for different values
of $b$, and summarized the results (which include the original value of $b$) in
Table~\protect\ref{F_over_I}.

\begin{table}[htbp!]
\caption{Relation $F/I$ for different distances between electrodes}
\centering
\begin{tabular}{| l | l | l | }
\hline\hline
   Distance relative to original &  Calculated $F/I$ &  $F/I$ approximated by $1.284 b/\mu$ \\
\hline
   0.5   &    89.42  &  89.88      \\
   0.75  &    134.41 &  134.82     \\
   1     &    179.72 &  179.76     \\
   1.5   &    269.39 &  269.64     \\
   2     &    358.82 &  359.5      \\
\hline  
\end{tabular}\\
\label{F_over_I}
\end{table}

Obviously, one can express this relation
with an excellent accuracy by:

\begin{equation}
F = \frac{I (1.284b)}{\mu}.
\label{foveri1}
\end{equation}

\begin{figure}[h]
{\par\centering \includegraphics{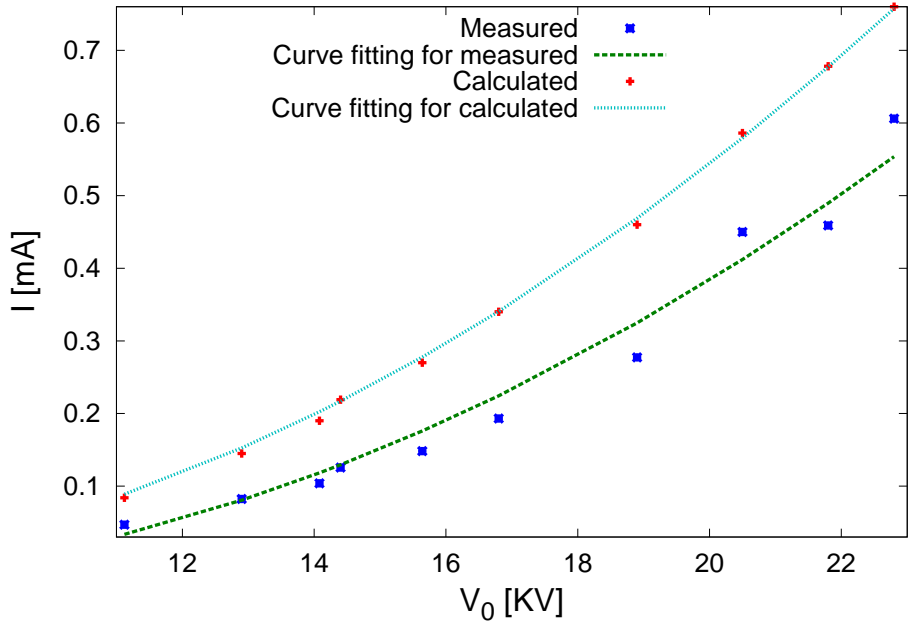} \par}
%Created from voltage_current_calc_meas.gif, width=150
\caption{(color online) The current-voltage curve of the lifter. The '*' describe the
measured results and the '+' describe the calculated results. The
curves are the closest approximations to the above results
using $I=K_1 V_0 (V_0 - \rm{CIV})$. For the calculated results
$K_1=2.16\, \rm{mA/KV}^2$ and $\rm{CIV}=7.42KV$, while for the measured results
$K_1=1.82\, \rm{mA/KV}^2$ and $\rm{CIV}=9.46KV$.}
\label{curr_volt}
\end{figure}

Now we analyze the current-voltage relation for the calculated and measured
results, which are shown in Figure~\protect\ref{curr_volt}.

Both measured and calculated results may be curve fitted
with the formula

\begin{equation}
I=K_1 V_0 (V_0 - \rm{CIV}),
\label{towns}
\end{equation}

although the measured results look farther from their fitted curve than the
calculated results (might be because of some technical problems during
experiment).

The curve fitted to the calculated results used $K_1=2.16\, \rm{mA/KV}^2$
and $\rm{CIV}=7.42KV$, while this fitted to the
measured results used $K_1=1.82\, \rm{mA/KV}^2$ and
$\rm{CIV}=9.46KV$.

We remark that the measured results show
a too high CIV and this is connected to the fact that the correspondence
between measured and calculated results is poor for low currents.
This might be explained by the fact that the CIV is affected by
additional factors like temperature, air pressure, or partial damage
to corona wire, because of repeated operation \protect\cite{peek}.
On the other hand, higher values of current and voltage are less
affected by the CIV (see eq.~(\protect\ref{towns})), and there
we get a much better correspondence between measured and calculated
current.

For a point to horizontal plane Warburg \protect\cite{warburg1} found the
constant $K_1$ to be proportional to the distance between the electrodes
to the power $-3.17$. This power has been corrected to $-3$ by Jones
\protect\cite{jones1}.

For concentric cylinders, Townsend \protect\cite{townsend} found the
constant $K_1$ to be proportional to the capacitance, and the distance
between electrodes at power $-2$, namely: $K_1=4\mu C/R^2$, where $\mu$
is the ion mobility and $R$ is the outer cylinder radius.

We would expect our case, being two dimensional, to behave similar to 
Townsend's concentric cylinders, i.e. $K_1$ should be some constant
multiplied by $\mu C/b^2$.
This assumption is tested with the results obtained for different values
of $b$, after curve fitting, and summarized in Table~\protect\ref{K1}.

\begin{table}[htbp!]
\caption{Constant $K_1$ of the relation $I=K_1 V_0 (V_0 - \rm{CIV})$ (in units of $\rm{mA/KV^2}$)
 for different distances between electrodes}
\centering
\begin{tabular}{| l | l | l | }
\hline\hline
   Distance relative to original &  Curve fitted $K_1$ &  $K_1$ approximated by $1.866 \mu C / b^2$ \\
\hline
   0.5   &    9.07   &  9.63      \\
   0.75  &    3.89   &  4.04      \\
   1     &    2.16   &  2.17      \\
   1.5   &    0.96   &  0.91      \\
   2     &    0.53   &  0.5       \\
\hline  
\end{tabular}\\
\label{K1}
\end{table}

We see that $K_1$ can be expressed by the approximate formula

\begin{equation}
K_1 = 1.866 \mu C / b^2
\label{cns}
\end{equation}

with an accuracy of about $5\%$.

We may combine the eqs.~(\protect\ref{foveri1}), (\protect\ref{towns}),
(\protect\ref{cns}), with eq.~(\protect\ref{approxVoverE}) and
(\protect\ref{peekformula}) to get an approximate formula for the thrust
on the lifter as function of the applied voltage $V_0$:

\begin{equation}
F=4.792 \pi\epsilon_0 l \frac{V_0}{b} \left(\frac{V_0}{1.011 \ln(b/a)+1.3035}-3\times 10^6 \left(a+.03 \sqrt{a}\right)\right),
\label{approx_force}\\
\end{equation}

where all the magnitudes are in MKS units.

We remark that the mobility $\mu$ canceled out. This does not mean
that eq.~(\protect\ref{approx_force}) works for negative ions, because
as explained in Sect.~2, the calculations done in this paper are
valid only for positive corona.
Also we see that the force increases as the distance $b$ decreases.\\

\noindent{\bf\large CONCLUSIONS}\\

In this paper we calculated the force on a levitation unit, built as a
thin anode wire over a vertical cathode plane, as the electric force
on the space charge and compared with experiment.

Our calculations were based on the Laplacian solution for this
electrostatic configuration, which by itself is a new result.

With the aid of the Laplacian solution and Deutsch assumption
we were able to calculate the force on the
levitation unit and the current. The calculated results showed a
good correspondence with the measured results.

Also, we derived for this type of lifter the relation between
thrust and current in eq.~(\protect\ref{foveri1}) and the
current-voltage characteristic curve in equations (\protect\ref{towns})
and (\protect\ref{cns}).

And finally based on the above relations, we derived an approximate
formula for the force as function of the applied voltage in
eq.~(\protect\ref{approx_force}).

\noindent

\clearpage

\renewcommand{\theequation}{A.\arabic{equation}} \setcounter{equation}{0}
\appendix

\section{The calculation of the Laplacian solution}

First we require condition (\protect\ref{boundary12dercont}):

\begin{equation}
\frac{-1}{\ln(b/a)}+\sum_{m=1}^{\infty}m A_m \left( (b/a)^m + (b/a)^{-m}\right) \cos(m\varphi)=\sum_{m=1}^{\infty}\frac{-m B_m}{2(1-\alpha/\pi)} \sin\left(\frac{m}{2}\frac{\varphi-\alpha}{1-\alpha/\pi}\right)
\label{boundary_der_pot_r_eq_b}
\end{equation}

and the above condition holds for $\alpha\le\varphi\le 2\pi-\alpha$. In this range, the
$\sin\left(\frac{m}{2}\frac{\varphi-\alpha}{1-\alpha/\pi}\right)$ functions are orthogonal, so we
multiply eq.~(\protect\ref{boundary_der_pot_r_eq_b}) by $\sin\left(\frac{n}{2}\frac{\varphi-\alpha}{1-\alpha/\pi}\right)$
(for any positive integer $n$) and integrate on $\varphi$ over the range
$[\alpha, 2\pi-\alpha]$. We use the following integrals:

\begin{equation}
\int_{\alpha}^{2\pi-\alpha}\sin\left(\frac{n}{2}\frac{\varphi-\alpha}{1-\alpha/\pi}\right)d\varphi=\left\{
\begin{array}{ll}
0 & n \; \mathrm{even} \\
\frac{4}{n}(1-\alpha/\pi) & n \; \mathrm{odd} \\
\end{array}
\right.
\label{int_sin}
\end{equation}

\begin{equation}
\int_{\alpha}^{2\pi-\alpha}\sin\left(\frac{n}{2}\frac{\varphi-\alpha}{1-\alpha/\pi}\right)\sin\left(\frac{m}{2}\frac{\varphi-\alpha}{1-\alpha/\pi}\right)d\varphi=\pi(1-\alpha/\pi)\delta_{mn},
\label{int_sin_sin}
\end{equation}

where $\delta_{mn}$ is the Kronecker delta. We also use:

\begin{equation}
\int_{\alpha}^{2\pi-\alpha}\sin\left(\frac{n}{2}\frac{\varphi-\alpha}{1-\alpha/\pi}\right)\cos(m\varphi)d\varphi=G_{nm},
\label{int_sin_cos}
\end{equation}

where $G_{nm}$ is a matrix defined by

\begin{equation}
G_{nm}=\left\{
\begin{array}{lll}
0 & & n \; \mathrm{even} \\
\cos(m\alpha)\frac{n/(1-\alpha/\pi)}{((n/2)/(1-\alpha/\pi))^2-m^2} & m\neq(n/2)/(1-\alpha/\pi) & n \; \mathrm{odd} \\
-(\pi-\alpha)\sin(m\alpha) & m=(n/2)/(1-\alpha/\pi) & n \; \mathrm{odd} \\
\end{array}
\right. .
\label{Gnm}
\end{equation}

The last case defined by $m=(n/2)(1-\alpha/\pi)$ is not likely to
happen if $\alpha\rightarrow 0$, except for very specific values of
$\alpha$, but we calculated
this case for completeness. It is to be mentioned that for this case
$\alpha/\pi=1-n/(2m)$, for some specific $m$ and $n$, hence
$m\alpha=m\pi-n\pi/2$ is equivalent to the points $\pi/2$ or $3\pi/2$ on
the unity circle, so that $\cos m\alpha=0$, and this has been used in
the above calculation. Also, $\sin(m\alpha)$ could be written as
$(-1)^{m-n/2-1/2}$. 

After performing the above integrals, we obtain from
eq.~(\protect\ref{boundary_der_pot_r_eq_b}) the following result for the
$B$ coefficients in eq.~(\protect\ref{expression2_V2}):

\begin{equation}
-\frac{1}{2}\pi n B_n=\left\{
\begin{array}{ll}
0 & n \; \mathrm{even} \\
-\frac{4}{n}(1-\alpha/\pi)\frac{1}{\ln(b/a)}+\sum_{m=1}^{\infty}G_{nm} m A_m \left( (b/a)^m + (b/a)^{-m}\right) & n \; \mathrm{odd} \\
\end{array}
\right. .
\label{Bn}
\end{equation}

This result implies that $B_n$ are $0$ for even $n$, and this is expected
because of the mirror symmetry around the $x$ axis. 

Now we require conditions (\protect\ref{pot_cont_b}):

\begin{equation}
{\scriptstyle L_0+\sum_{m=1}^{\infty} A_m \left( (b/a)^m - (b/a)^{-m}\right) \cos(m\varphi) = } \left\{
\begin{array}{ll}
0 & 0\le\varphi\le\alpha \\
0 & 2\pi-\alpha\le\varphi\le 2\pi \\
{\scriptstyle \sum_{m=1}^{\infty} B_m \sin\left(\frac{m}{2}\frac{\varphi-\alpha}{1-\alpha/\pi}\right) } & \alpha\le\varphi\le 2\pi-\alpha \\
\end{array}
\right. .
\label{boundary_pot_r_eq_b}
\end{equation}

The above condition is defined for $0\le\varphi\le 2\pi$. In this
range, the $\cos(m\varphi)$ functions are orthogonal, so we multiply
eq.~(\protect\ref{boundary_pot_r_eq_b}) by $\cos(n\varphi)$ and integrate
on $\varphi$ over the range $[0, 2\pi]$, for $n$ being any non
negative integer.

First, if we do it for $n=0$, the sum on the left side vanishes, and
using eq.~(\protect\ref{int_sin}), with $m$ instead of $n$, we obtain the
following expression for $L_0$:

\begin{equation}
L_0=\frac{2}{\pi}(1-\alpha/\pi)\sum_{m=1 \; \mathrm{(odd)}}^{\infty} \frac{B_m}{m},
\label{expression1_L0}
\end{equation}

and we could have even omitted the word ``odd'', because we already know
that the even $B$ coefficients from eq.~(\protect\ref{expression2_V2}) are $0$.

Now we multiply eq.~(\protect\ref{boundary_pot_r_eq_b}) by
$\cos(n\varphi)$ and integrate from $\varphi=0$ to $2\pi$, for $n\neq 0$.
This time $L_0$ vanishes, and after using
$\int_0^{2\pi}\cos(n\varphi)\cos(m\varphi)d\varphi=\pi\delta_{mn}$,
and the result
(\protect\ref{int_sin_cos}) for $m$ and $n$ switched, we obtain
from eq.~(\protect\ref{boundary_pot_r_eq_b}) the following result for
the $A$ coefficients in eq.~(\protect\ref{expression2_V1}):

\begin{equation}
\pi A_n \left( (b/a)^n  - (b/a)^{-n}\right) = \sum_{m=1 \; \mathrm{(odd)}}^{\infty} G_{mn} B_m.
\label{An}
\end{equation}

We redefine now the odd indexes (like $m$ in eq.~\protect\ref{An}) by $2l-1$,
$l$ going from $1$ to $\infty$. Hence the vector $B_m$ and the matrix
$G_{mn}$ (like in eq.~\protect\ref{An}) are redefined to $B_l$ and $G_{lm}$.

This applies to equations (\protect\ref{expression2_V2}), 
(\protect\ref{Bn}), (\protect\ref{expression1_L0}) and (\protect\ref{An}).

We rewrite now eq.~(\protect\ref{Bn}) and (\protect\ref{An})
in matrix form, to solve 2 matrix equations with 2 unknown vectors for the
coefficients $A$ and $B$, then find $L_0$ with eq.~(\protect\ref{expression1_L0})
and eventually calculate the potential and scale it by the factor
$V_0/(1+L_0)$.

Hence we define the
diagonal matrix $\mathcal{Q}$, by its components as
$Q_{mn}\equiv \left( (b/a)^m - (b/a)^{-m}\right)\delta_{mn}$, and obtain:

\begin{equation}
\pi \mathcal{Q} \overline{A} =  \mathcal{G}^T \overline{B}
\label{expr3_A}
\end{equation}

and we define the
diagonal matrices $\mathcal{P}$, by 
$P_{mn}\equiv \left( (b/a)^m + (b/a)^{-m}\right)\delta_{mn}$, 
$\mathcal{S}$, by $S_{mn}\equiv (2m-1)\delta_{mn}$ and 
$\mathcal{J}$, by $J_{mn}\equiv m\delta_{mn}$. We also define the
column vector $\overline{K}$ by its components $K_m=1$, obtaining:

\begin{equation}
-\frac{1}{2}\pi\mathcal{S}\overline{B}=-4(1-\alpha/\pi)\frac{1}{\ln(b/a)}\mathcal{S}^{-1}\overline{K}+\mathcal{G}\mathcal{J}\mathcal{P}\overline{A}.
\label{expr3_B}
\end{equation}

We isolate $\overline{B}$ from eq.~(\protect\ref{expr3_B}):

\begin{equation}
\overline{B}=\frac{8}{\pi}(1-\alpha/\pi)\frac{1}{\ln(b/a)}\mathcal{S}^{-2}\overline{K}-\frac{2}{\pi}\mathcal{S}^{-1}\mathcal{G}\mathcal{J}\mathcal{P}\overline{A} 
\label{expr4_B}
\end{equation}

and set it in eq.~(\protect\ref{expr3_A}) to obtain a closed form solution
for $\overline{A}$:

\begin{equation}
\overline{A}=\frac{8}{\pi}(1-\alpha/\pi)\frac{1}{\ln(b/a)}(\pi\mathcal{Q}+\frac{2}{\pi}\mathcal{G}^T \mathcal{S}^{-1}\mathcal{G}\mathcal{J}\mathcal{P})^{-1}\mathcal{G}^T\mathcal{S}^{-2}\overline{K}.
\label{expr4_A}
\end{equation}

Clearly, for $\alpha=\pi$ (concentric cylinders), $\overline{A}$,
$\overline{B}$ and $L_0$ are all $0$, hence $V_2=0$ and we are left
only with the trivial solution for $V_1$, as expected.

%%\end{spacing}

\newpage

%\renewcommand{\thetable}{\arabic{table}} \setcounter{table}{0}
%{\large Tables} \vspace{1cm}

\clearpage

%\renewcommand{\thefigure}{\arabic{figure}} \setcounter{figure}{0}
%{\large Figure captions} \vspace{1cm}


\newpage

\begin{thebibliography}{15}
\bibitem{ttbrown1} T.T. Brown, ``A method of and an apparatus or machine for producing force or motion'' , British Patent 300311, (1928)

\bibitem{ttbrown2} T.T. Brown, ``Electrostatic motor'', US Patent 1974483, (1934)

\bibitem{ttbrown3} T.T. Brown, ``Electrokinetic Apparatus'', US Patent 2949550, (1960)

\bibitem{bb-web} Website dedicated to BB effect ``http://www.biefeldbrown.com/''
\bibitem{jln} Jean-Louis Naudin website ``http://jnaudin.free.fr/''

\bibitem{tajmar-matos} Tajmar M. and de Matos, C.J.,''Coupling of electromagnetism and gravitation in the weak
field approximation'', {\it Journal of Theoretics}, {\bf Vol.3} (1),  (Feb/March 2001).

\bibitem{takaaki-musha} Musha T., ``Theoretical explanation of the Biefeld-Brown Effect'', {\it Electric Space Craft Journal}, {\bf Issue 31} (2000).

\bibitem{montalk} Website ``http://montalk.net/science/84/the-biefeld-brown-effect''

\bibitem{bahder-fazi} T.B. Bahder, C. Fazi, ``Force on an asymmetrical capacitor, {\it Army Research Laboratory}, Report No. ARL-TR-3005, (March 2003) 

\bibitem{tajmar} Tajmar M. ``Biefeld-Brown Effect: Misinterpretation of Corona Wind Phenomena'',
{\it AIAA Journal}, {\bf Vol.42} (2), 315-318 (2004).

\bibitem{zhao-adamiak} L. Zhao, K. Adamiak, ``EHD gas flow in electrostatic levitation unit'', {\it Journal of Electrostatics}, {\bf Vol. 64}, 639-645 (July 2006)

\bibitem{zhao-adamiak1} L. Zhao, K. Adamiak, ``Numerical analysis of forces in an electrostatic levitation unit'', {\it Journal of Electrostatics}, {\bf Vol. 63}, 729-734 (June 2005)

\bibitem{blazelabs} Website ``http://www.blazelabs.com/''

\bibitem{deutsch} W. Deutsch, ``Uber die Dichteverteilung unipolarer Ionenstrome'', {\it Annalen der Physik}, {\bf Vol. 16}, 588-612 (1933)

\bibitem{tsyrlin} L. E. Tsyrlin {\it Sov. Phys.-Tech. Phys.} {\bf Vol. 30} (1958)

\bibitem{ieta} A. Ieta, Z. Kucerovsky and W. D. Greason, ``Laplacian approximation of Warburg distribution'' {\it Journal of Electrostatics}, {\bf Vol. 63 (2)}, 143-154 (February 2005) 

\bibitem{sigmond} R.S. Sigmond, ``Simple approximate treatment of unipolar space-charge-dominated coronas: the Warburg law and the saturation current'', {\it J. Appl. Phys.}, {\bf Vol. 53 (2)}, 891-898 (1982)

\bibitem{sigmond1} R.S. Sigmond, ``The unipolar corona space charge flow problem'', {\it Journal of Electrostatics}, {\bf Vol. 18}, 249-272 (1986)

\bibitem{amoruso} V. Amoruso and F. Lattarulo, ``Deutsch hypothesis revisited'', {\it Journal of Electrostatics}, {\bf Vol. 63}, 717-721 (2005)

\bibitem{jones} J.E. Jones and M. Davies, ``A critique of the Deutsch assumption'', {\it J. Phys. D: Appl. Phys.}, {\bf Vol. 25 (12)}, 1749-1759 (1992)

\bibitem{bouziane} A. Bouziane, K. Hidaka, J.E. Jones, A.R. Rowlands,
 M.C. Taplamacioglu, R.T. Waters, ``Paraxial corona discharge Part 2:
 Simulation and analysis'', {\it IEE Proc.-Sci. Meas. Technol.}, {\bf Vol. 141 (3)}, 205-214 (1994)

\bibitem{popkov} V.I. Popkov, ``On the theory of unipolar DC corona'', {\it Electrichestvo},
{\it Vol. 1}, 33-48 (1949)

\bibitem{sarmajan} M.P. Sarma and W. Janischewskyj, ``Analysis of corona losses on DC
transmission lines. Part I unipolar lines'', {\it IEEE Trans. Power Appar. Sysr.},
{\bf Vol. 88 (5)}, 718-731 1969

\bibitem{bouziane1} A. Bouziane, K. Hidaka, M.C. Taplamacioglu and R.T. Waters
``Assessment of corona models based on the Deutsch approximation'',
 {\it J. Phys. D: Appl. Phys.}, {\bf Vol. 27}, 320-329 (1994)

\bibitem{chen} J. Chen and J.H. Davidson, {\it Plasma Chemistry and Plasma Processing},
 {\it Vol. 23 (1)}, 83-102 (2003)

\bibitem{warburg1} E. Warburg, ``Uber die spitzenentladung'', {\it Wied. Ann}, {\bf Vol. 67} 69-83 (1899).

\bibitem{warburg2} E. Warburg, ``Charakteristik des spitzenstromes'', {\it Handbuch der Physik (Springer, Berlin)},
{\bf Vol. 14}, 154-155 (1927).

\bibitem{peek} F.W. Peek ``Dielectric Phenomena in High Voltage Engineering'',  {\bf McGraw-Hill} (1929)

\bibitem{felici} N.J. Felici, ``Recent advances in the analysis of D.C. ionized electric fields'', {\it Direct Current}, {\bf Vol. 8 (10)}, 278-287 (1963) 

\bibitem{goldman} M. Davies, A. Goldman, M. Goldman and J.S. Jones ``Developments in the theory of corona corrosion for negative coronas in air'', {\it Proc. XVIII Int. Conf. on Phenomena in Ionized Gases (ICPIG)}, (1987)

\bibitem{kaptzov} N. Kaptzov, ``Elektricheskie Yavlenia v Gazakh I Vakuumme'', {\bf Ogiz, Moscow}, 587-630 (1947)

\bibitem{feng} J.Q. Feng, ``An analysis of corona currents between two concentric cylindrical electrodes'', {\it Journal of Electrostatics}, {\bf Vol. 46}, 37-48 (1998)

\bibitem{townsend} J. S. Townsend, {\it Philos. Mag.}, {\bf Vol. 28}, p83 (1914).

\bibitem{kondo} Y. Kondo and Y. Miyoshi, {\it Jpn. J. Appl. Phys.}, {\bf Vol. 17}, 643 (1978).

\bibitem{selim1} A. Goldman, E. O. Selim, and R. T. Waters,  {\it The 5th International Conference on Gas Discharges, lEE Conf. Publ.}, {\bf No. 165} pp. 88-91, (London, 1978).

\bibitem{selim2} E. O. Selim and R. T. Waters, {\it Proceedings of the 3rd International Symposium on High Voltage Engineering}, Paper 53.03, (Milan, 1979)

\bibitem{selim3} E. O. Selim and R. T. Waters, {\it The 6th International Conference on Gas Discharges and their Applications, lEE Conf. Publ.},  {\bf No. 189}, pp. 146-149, (London, 1980)

\bibitem{jones1} J. E. Jones, ``A Theoretical Explanation of the Laws of Warburg and Sigmond'', {\it Proc. Roy. Soc. London A}, {\bf Vol. 453 (1960)}, pp. 1033-1052 (1997)

\end{thebibliography}
\end{document}